\documentclass[review]{elsarticle}

\usepackage{lineno,hyperref}
\modulolinenumbers[5]
\usepackage{graphicx,color}
\usepackage{wrapfig}
\usepackage{dcolumn}
\usepackage{gensymb}

\journal{ArXiv}

\begin{document}

\begin{frontmatter}

\title{Analyzing the magnetic influence on magneto-optical interactions}

\author{Francisco Estrada$^{1}$ and José Holanda$^{2, }$\corref{mycorrespondingauthor}}
\cortext[mycorrespondingauthor]{Corresponding author: joseholanda.silvajunior@ufrpe.br}
\address{$^{1}$Facultad de Biología, Universidad Michoacana de San Nicolas de Hidalgo, Av. F. J. Mujica s/n Cd. Universitaria, Morelia, Michoacán, México.}
\address{$^{2}$Programa de Pós-Graduação em Engenharia Física, Universidade Federal Rural de Pernambuco, 54518-430, Cabo de Santo Agostinho, Pernambuco, Brazil.}

\begin{abstract}
Here, we study the magneto-optical interactions in magnetic structures considering the dependence of the interactions with the magnetic field. We perform numerical simulations in a structure of magnetic nanowires, considering them as one chain of strongly interacting single-domain particles. Robustly, we obtain a quantitative value for the interactions, which allows us to classify them into two magnetic states: demagnetized and magnetized.
\end{abstract}

\begin{keyword}
\texttt{}Mgneto-optical\sep magnetic states\sep interacting system\sep energy balance
\end{keyword}
\end{frontmatter}
\vspace{1.5cm}

\section{Introduction}

The recent interest of the scientific community in magneto-optical interactions has opened the possibility of a global understanding of characteristics not yet dazzled in nanomaterials [1-6]. Minimizing the bit size in one magneto-optical system for data recording is a challenge for optoelectronics and spintronics applications [3-14]. Such reduction produces an increase in the interactions between components of the structure. The scientific interest actual is the quantification (codification) of these interactions. One of the systems that have a high particle distribution density is the arrays of magnetic nanowires electrodeposited in alumina membranes [6, 8-14]. This type of system can present different magnetization reversal modes with a predominant coherent configuration [2, 8-20]. Such systems can be strongly influenced by magneto-optical interactions [2-21]. The influence is detected mainly during the magnetization process with light, which always presents reversible and irreversible components. Furthermore, a striking feature of the magnetization process is that it is not possible to separate the parts of their hysteresis without losing information due to changes in the magnetic energies of the structure. This means that many properties remain hidden during the magnetization process and there is a need for understanding. The study from the magneto-optical interactions can be performed using the remanent state obtained during the magnetization process [2].  

In a particular system, the well-established normalized $\Delta m$ curves ($\Delta m_{N}$) produce results of the interaction effects, such curves are comparisons between isothermal remanent magnetization ($IRM(H)$) and direct current demagnetization ($DCD(H)$) curves, which define other physical quantities such as $m_{d}(H) =$ $DCD(H)/IRM(H_{Max})$ and $m_{r}(H) =$ $IRM(H)/IRM(H_{Max})$ that are normalized considering the value obtained with maximum magnetic field [2, 17, 18, 22-27]. In this paper we present a numerical study on the predominant magneto-optical interactions in structures, for that, we perform numerical simulations in a system of magnetic nanowires, considering them as one chain of strongly interacting single-domain particles. After analyzing magneto-optical interactions, we observe two types of magnetic states, i. e., magnetized and demagnetized, which reveal the main characteristics of magneto-optical energies. Our approach seeks to describe the light-matter interaction with the application of a magnetic field, where the light only serves to excite the magneto-optical effects, that is, all results are obtained considering the dependence of magneto-optical interactions with the magnetic field.

\section{Continuous approach}

The effects of the magnetic interactions in structures have been studied by using $\Delta m$ curves [17, 18, 22-29] or discrete models without analyzing the dependence directly on the magnetic field and light [2]. The $\Delta m$ curves are obtained through the relationships between the $m_{d}(H)$ and $m_{r}(H)$ curves, where the initial magnetic state of the structure differentiates them. The model proposed by Stoner Wohlfarth [17, 27] reveals an intrinsic relationship between $m_{d}(H)$ and $m_{r}(H)$ for non-interacting structures. Based on this fact, we propose here that the magneto-optical interactions for non-interacting particles have one associated intensity that can be written as
\begin{equation} 
	I_{N-I} = \left|\int_{H_{i}}^{H_{f}}\left( \frac{\eta_{N-I}}{\Delta H}\right)dH\right|,
	\label{1}
\end{equation}
where $H_{f} > H_{i}$ and $\eta_{N-I} = \left[1-2m_{r}(H)\right]-m_{d}(H)$. The magnetic fields $H_f$ and $H_i$ represent the maximum and minimum fields in the interval $\Delta H = H_f - H_i$, respectively. In most experimental systems, the minimum magnetic field $H_i$ is zero. To better describe real systems, Henkel postulated that the difference in this behavior in a simple design was due to the interactions between the part of the structure. Usually, in experimental measurements, the data are far from the curve obtained with equation (1) [17, 18, 22-29]. Thus, equation (1) considers that the magnetization and demagnetization processes are the same. Qualitatively, the type of interaction was defined by inserting a term $\Delta m$ in equation (1). Based on this, we propose here that the intensity of the magneto-optical interactions for interacting particles can be written
\begin{equation} 
	I_{I} =  \left|\int_{H_{i}}^{H_{f}}\left(\frac{\eta_{I}}{\Delta H}\right)dH\right|,
	\label{2}
\end{equation}
where $\eta_{I} = \Delta m_{N} +  \left[1-2m_{r}(H)\right]-m_{d}(H)$. The indices $N-I$ and $I$ in equations (1) and (2) are associated with $\eta$ for systems without and with interactions. When the predominant magneto-optical interactions are demagnetizing (PMOID), $\Delta m_{N} <$ 0; and when the predominant magneto-optical interactions are magnetizing (PMOIM), $\Delta m_{N} >$ 0. The methodology presented here for the calculation from the magneto-optical interactions uses equations (1) and (2) resulting in a numerical value for the magneto-optical interactions of the structure. Hence, the intensity value from the magneto-optical interactions is obtained through Eqs. (1) and (2) by
\begin{equation} 
	I =  \left|\int_{H_{i}}^{H_{f}}\left( \frac{\Delta m_{N}}{\Delta H}\right)dH\right|.
	\label{3}
\end{equation}
Equation (3) represents both PMOID and PMOIM interactions so that we can rewrite it as
\begin{equation} 
	I_{k} =  \left|\int_{H_{i}}^{H_{f}}\left( \frac{\Delta m_{N}}{\Delta H}\right)dH\right| \longrightarrow 
	\left\{
	\begin{array}{ll}
		\Delta m_{N} = \Delta m^{D}/|\Delta m^{D}_{Max}| < 0 &$for PMOID$, \\
		\Delta m_{N} = \Delta m^{M}/|\Delta m^{M}_{Max}| > 0 &$for PMOIM$.
	\end{array}
	\right.
	\label{4}
\end{equation}
where \textit{k} = PMOID ($\Delta m_{N} < 0$) or PMOIM ($\Delta m_{N} > 0$).

\section{Results and discussion}

Our results were obtained considering a nanowire as one chain of interacting ellipsoidal grains according to Figure 1. 
\begin{figure}[h]
	\vspace{0.1mm} \hspace{0.1mm}
	\begin{center}
		\includegraphics[scale=0.4]{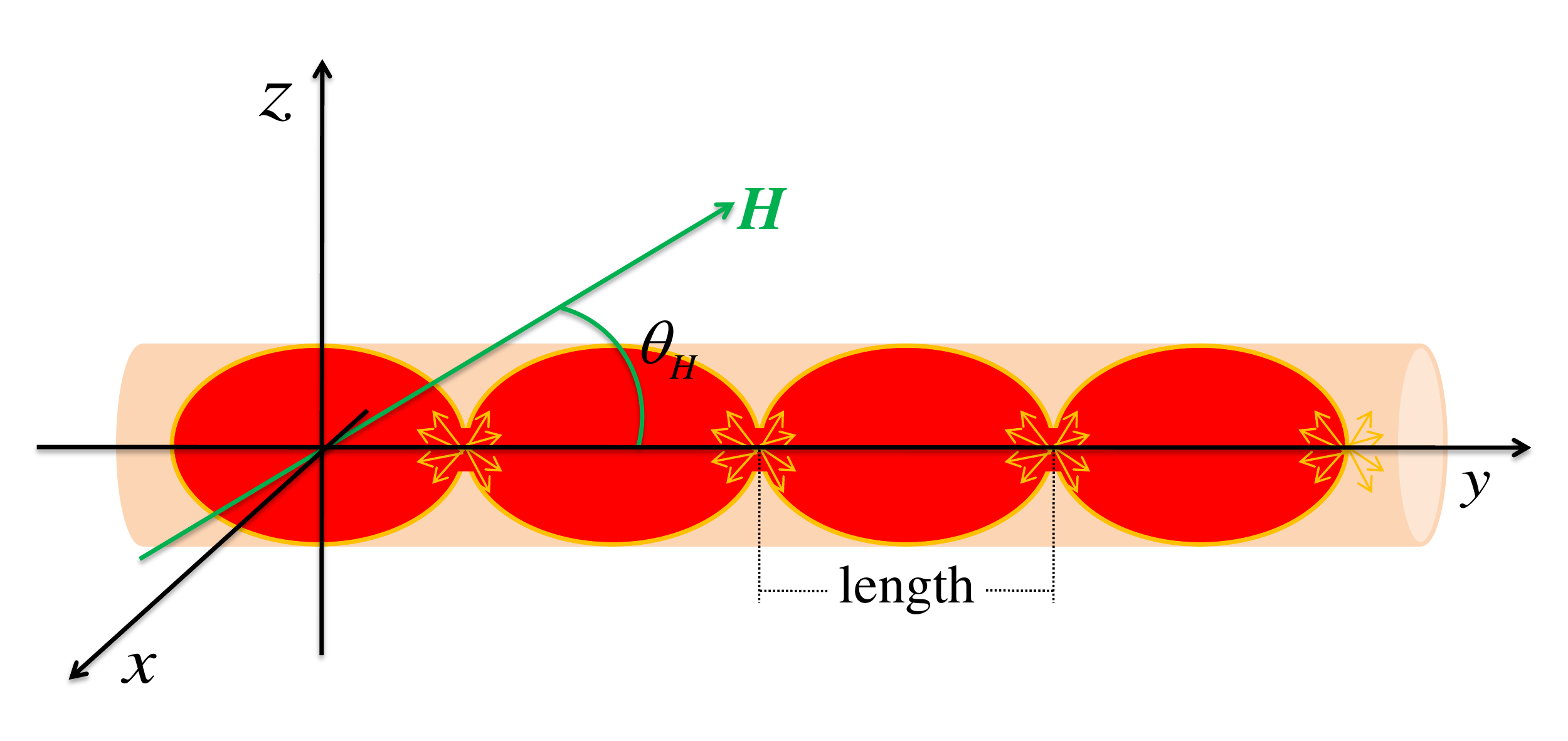}
		\caption{\label{arttype}(Color online) Schematic of a nanowire (length much larger than the diameter) as one chain of interacting single-domain particles, which are ellipsoidal grains.}
		\label{puga}
	\end{center}
\end{figure}
Experimentally, the length of the grains is on the order of 20 to 100 nm, as in single-domain particles [3-14]. Demagnetizing interactions are obtained frequently in experimental measurements [17, 18, 22-29], this is due to, in most cases, the magnetic response of structures being contrary to the applied magnetic field $H$, leaving the system in a global demagnetized state. Usually, the dependence of $\Delta m_{N}$ in terms of applied magnetic field $H$ on systems involving PMOID behavior is $\Delta m^{D} / |\Delta m_{Max}^{D}|$, where $\Delta  m^{D} = -\left(H_{C}^{D}\right)^2/\left[(H-H_{C}^{D})^2+(\Delta J_{D})^2\right]$, $H_{C}^{D}$ is the critical demagnetizing field (for $|\Delta m^{D}|$ maximum, i.e., for $|\Delta m_{Max}^{D}|$) and $\Delta J_{D}$ is the demagnetizing linewidth. For all our calculations, we considered $\Delta J_{D} = 1\% H_{C}^{D}$. We did this to condense the energy balance of the magneto-optical interactions so that they would produce a kind of avalanche. Figure 2 shows the interactions (Regions I) considering the $\Delta m_{N}$ curves obtained with $H_{C}^{D}$ = 0.75 kOe (Figure 2(a)), $H_{C}^{D}$ = 2.5 Oke (Figure 2(b)), $H_{C}^{D}$ = 5.0 kOe (Figure 2(c)) and $H_{C}^{D}$ = 7.5 kOe (Figure 2(d)), 
\begin{figure}[h]
	\vspace{0.1mm} \hspace{0.1mm}
	\begin{center}
		\includegraphics[scale=0.38]{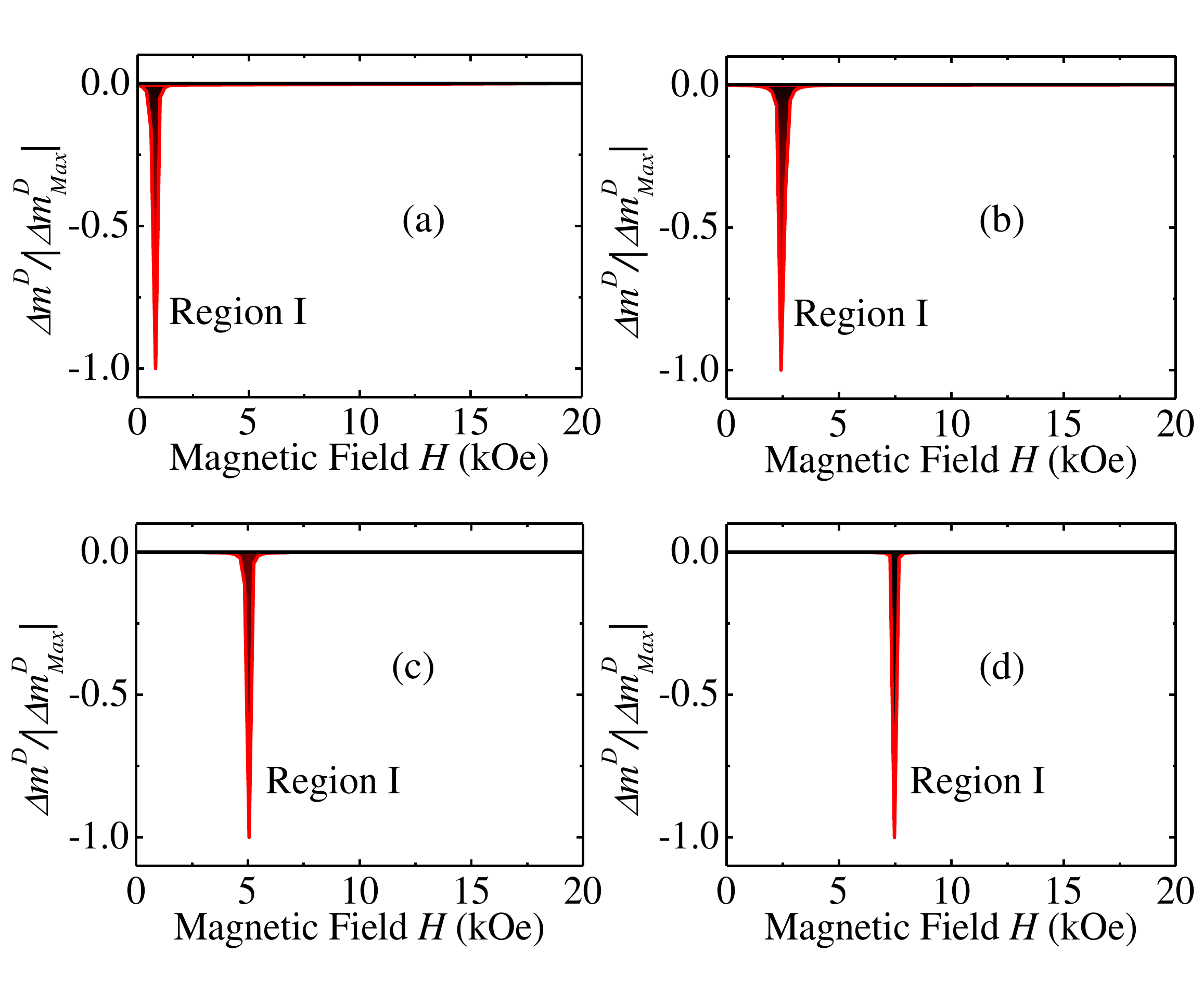}
		\caption{\label{arttype}(Color online) Interaction maps calculated from the magneto-optical interactions in terms of the applied magnetic field to different critical demagnetizing field $H_{C}^{D}$. (a) $H_{C}^{D}$ = 0.75 kOe, (b) $H_{C}^{D}$ = 2.5 Oke , (c) $H_{C}^{D}$ = 5.0 kOe and (d) $H_{C}^{D}$ = 7.5 kOe. The interaction maps were results obtained with the equations (1) and (2), and the intensities were calculated using the equation (4), which are $I_{PMOID}$ = 0.26 (item a), $I_{PMOID}$ = 0.32 (item b), $I_{PMOID}$ = 0.25 (item c), and $I_{PMOID}$ = 0.21 (item d).}
		\label{puga}
	\end{center}
\end{figure}
which are defined as interaction maps. The intensities values of the interactions were numerically calculated using the Eq. (4), which are $I_{PMOID}$ = 0.26, 0.32, 0.25, and 0.21, for Figures 2 (a), (b), (c), and (d), respectively. In this system, the magnetic behavior can often create magnetizing effects due to the exchange interactions of the structure, which are extremely important for the excitation of spin waves [30-34]. Such exchange interactions arise due to defects between grains that cause propagation of domain walls (see insert of Figure 3(a)).
\begin{figure}[h]
	\vspace{0.1mm} \hspace{0.1mm}
	\begin{center}
		\includegraphics[scale=0.38]{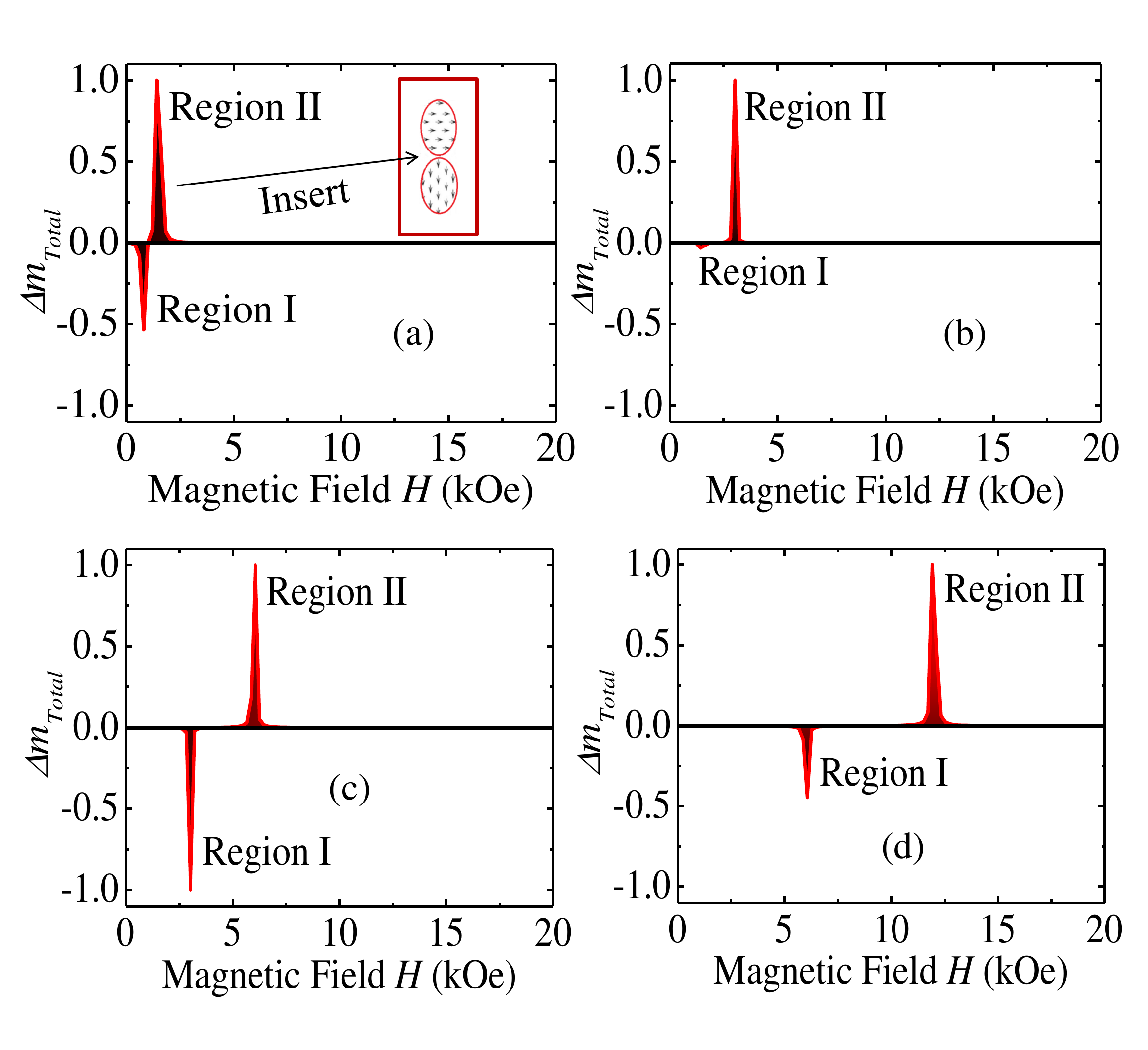}
		\caption{\label{arttype}(Color online) Interaction maps calculated from the magnetic interactions in terms of the applied magnetic field to different critical fields demagnetizing $H_{C}^{D}$ and magnetizing $H_{C}^{M}$. (a) $H_{C}^{D}$ = 0.75 kOe and $H_{C}^{M}$ = 1.5 kOe, (b) $H_{C}^{D}$ = 1.5 kOe and $H_{C}^{M}$ = 3.0 kOe, (c) $H_{C}^{D}$ = 3.0 kOe and $H_{C}^{M}$ = 6.0 kOe, and (d) $H_{C}^{D}$ = 6.0 kOe and $H_{C}^{M}$ = 12 kOe. The interaction maps were results determined with the equations (1) and (2), and the intensities were calculated using the equation (4), which are for the PMOID interactions, $I_{PMOIM}$ = 0.12 (item a), $I_{PMOIM}$ = 0.01 (item b), $I_{PMOIM}$ = 0.21 (item c), and $I_{PMOIM}$ = 0.12 (item d) and for PMOIM interactions, $I_{PMOIM}$ = 0.35 (item a), $I_{PMOIM}$ = 0.21 (item b), $I_{PMOIM}$ = 0.27 (item c), and $I_{PMOIM}$ = 0.35 (item d).}
		\label{puga}
	\end{center}
\end{figure}

According to the spin-wave theory [30, 31], effects that produce exchange interactions are relevant for their study and understanding. The proposal that we present can analyze the global knowledge of such results. To analyze the possible PMOIM behavior, we consider the effects that describe them (PMOID and PMOIM) as, $\Delta m_{Total} = \Delta m^{D}/|\Delta m^{D}_{Max}| + \Delta m^{M}/|\Delta m^{M}_{Max}|$, where $\Delta m^{D} =   -\left(H_{C}^{D}\right)^2/\left[(H-H_{C}^{D})^2+(\Delta J_{D})^2\right]$ and $\Delta m^{M} =  \left(H_{C}^{M}\right)^2/\left[(H-H_{C}^{M})^2\right.$
$\left.+(\Delta J_{M})^2\right]$, which can also describe very well experimental measurements obtained in the laboratory [17, 18, 22-29]. Here, $\Delta J_{M}$ is the magnetizing linewidth, in which we consider $\Delta J_{M}$ = $1\% H_{C}^{M}$ for all calculations. Figures 3 (a), (b), (c), and (d) show the PMOID behavior (Regions I) for the different $\Delta m^{D}/|\Delta m^{D}_{Max}|$ curves obtained with critical magnetic fields of $H_{C}^{D}$ = 0.75 kOe, $H_{C}^{D}$ = 1.5 kOe, $H_{C}^{D}$ = 3.0 kOe, and $H_{C}^{D}$ = 6.0 kOe, respectively. The same Figures 3 (a), (b), (c), and (d) also show the PMOIM behavior (Regions II) for the different $\Delta m^{M}/|\Delta m^{M}_{Max}|$ curves obtained with critical fields of $H_{C}^{M}$ = 1.5 kOe, $H_{C}^{M}$ = 3.0 kOe, $H_{C}^{M}$ = 6.0 kOe, and $H_{C}^{M}$ = 12 kOe, respectively. As results obtained with the calculation of PMOID and PMOIM interaction maps of Figure 3 using the Eq. (4), we bring the values, $I_{PMOID}$ = 0.12, 0.01, 0.21, and 0.12 for PMOID interactions, and $I_{PMOIM}$ = 0.35, 0.21, 0.27, and 0.35 for PMOIM interactions, which are shown in Table 1. 
\begin{table}[h]
	\caption{\label{tab:table4}{Shows the intensity values of interactions (PMOID and PMOIM) obtained from the interaction maps of Figure 3.}}%
	\begin{center}
		\begin{tabular}{cccc}
			$H_{C}^{D}$ (kOe) & $H_{C}^{M}$ (kOe) & $I_{PMOID}$ & $I_{PMOIM}$ \\
			\hline
			0.75 & 1.5 & 0.12 & 0.35 \\
			1.5 & 3.0 & 0.01 & 0.21  \\
			3.0 & 6.0 & 0.21 & 0.27 \\
			6.0 & 12 & 0.12 & 0.35 \\
		\end{tabular} 
	\end{center}
\end{table}

The demagnetizing and magnetizing effects are results of the energy balance due to the magnetization process as also shown in Table 1. To understand the angular behavior of magneto-optical interactions in a strongly interacting system, we present in Figure 4 the angular dependence from the interactions 
\begin{figure}[h]
	\vspace{0.1mm} \hspace{0.1mm}
	\begin{center}
		\includegraphics[scale=0.38]{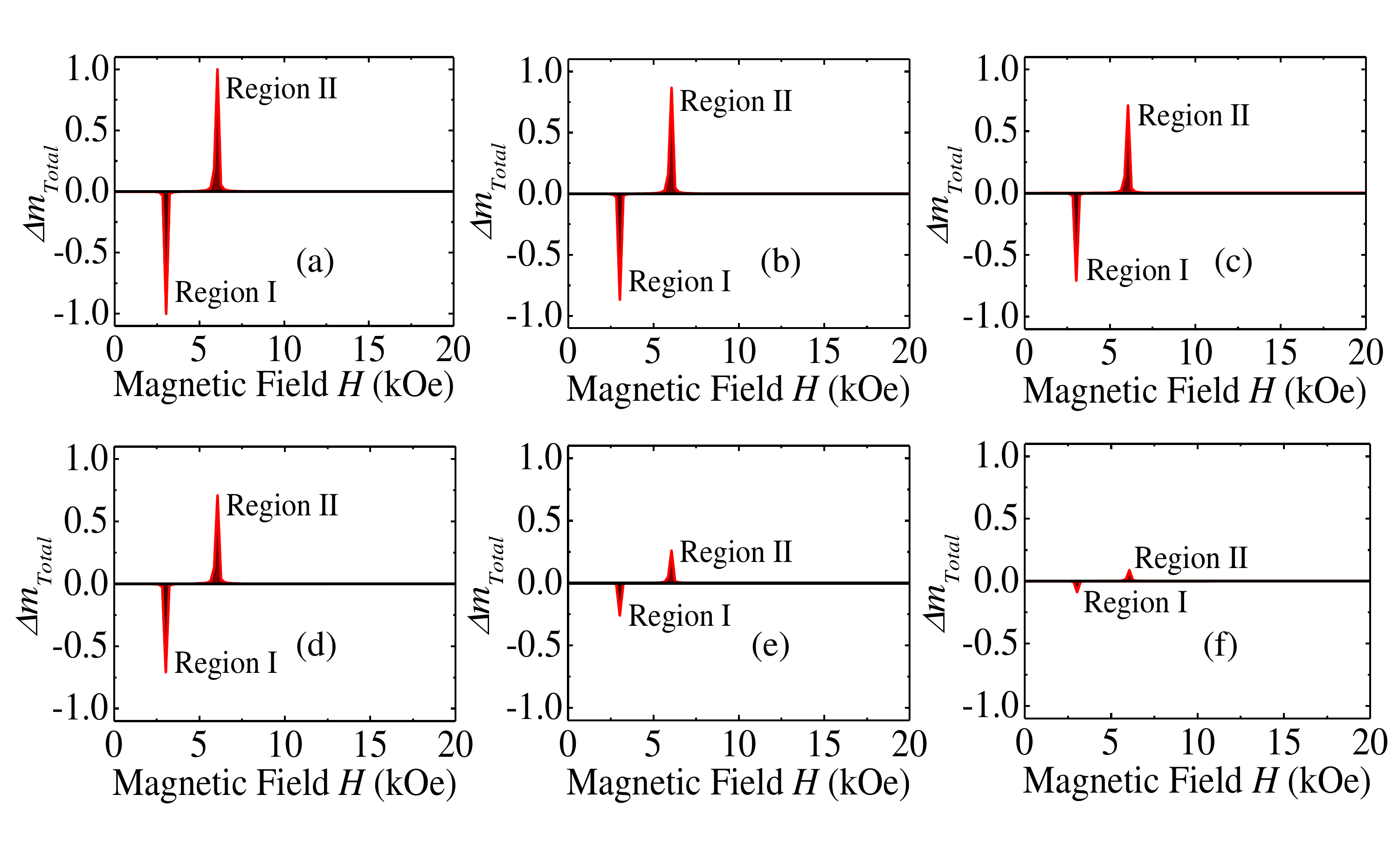}
		\caption{\label{arttype}(Color online) Angular dependence of magneto-optical interactions for different angles. The interaction maps were results obtained with the equations (1), (2), and $\Delta m_{Total} = \left(\Delta m^{D}/|\Delta m^{D}_{Max}| + \Delta m^{M}/|\Delta m^{M}_{Max}|\right)cos(\theta_{H})$ to $H_{C}^{D}$ = 3 kOe and $H_{C}^{M}$ = 6 kOe. (a) $\theta_{H}$ = 0°, (b) $\theta_{H}$ = 30°, (c) $\theta_{H}$ = 45°, (d) $\theta_{H}$ = 60°, (e) $\theta_{H}$ = 75° and (f) $\theta_{H}$ = 85°.}
		\label{puga}
	\end{center}
\end{figure}
considering the relation $\Delta m_{Total} = \left(\Delta m^{D}/|\Delta m^{D}_{Max}| + \Delta m^{M}/|\Delta m^{M}_{Max}|\right)$ $cos(\theta_{H})$ [2, 8, 13, 16, 29] to $H_{C}^{D}$ = 3 kOe and $H_{C}^{M}$ = 6 kOe. TThis approach describes in detail the behavior of magneto-optical interactions in a magnetic nanowire modeled as a chain of interacting ellipsoidal grains [8, 25-28]. Such dependence shows that for a magnetic field applied parallel to the wire axis, the PMOID and PMOIM behaviors are maximum and decrease with increasing angle $\theta_{H}$. This behavior is result of the decrease in interactions between the grains as the magnetic field becomes perpendicular to the wire axis. In Figure 5 we show the general variation of the intensity values from the magneto-optical interactions ($I_{PMOID}$ and $I_{PMOIM}$) as a function of the angle $\theta_{H}$, where we use $\Delta m_{Total} = \left(\Delta m^{D}/|\Delta m^{D}_{Max}| + \Delta m^{M}/|\Delta m^{M}_{Max}|\right)cos(\theta_{H})$ with $H_{C}^{D}$ = 3 kOe, $\Delta J_{D} = 1\% H_{C}^{D}$, $H_{C}^{M}$ = 6 kOe and $\Delta J_{M} = 1\% H_{C}^{M}$. We define for any wavelength the best conditions to observe the maximum number of magneto-optical interactions. We also observed that it is possible increase the intensity of interactions when the PMIOD and PMOIM effects have similar intensities.
\begin{figure}[h]
	\vspace{0.1mm} \hspace{0.1mm}
	\begin{center}
		\includegraphics[scale=0.28]{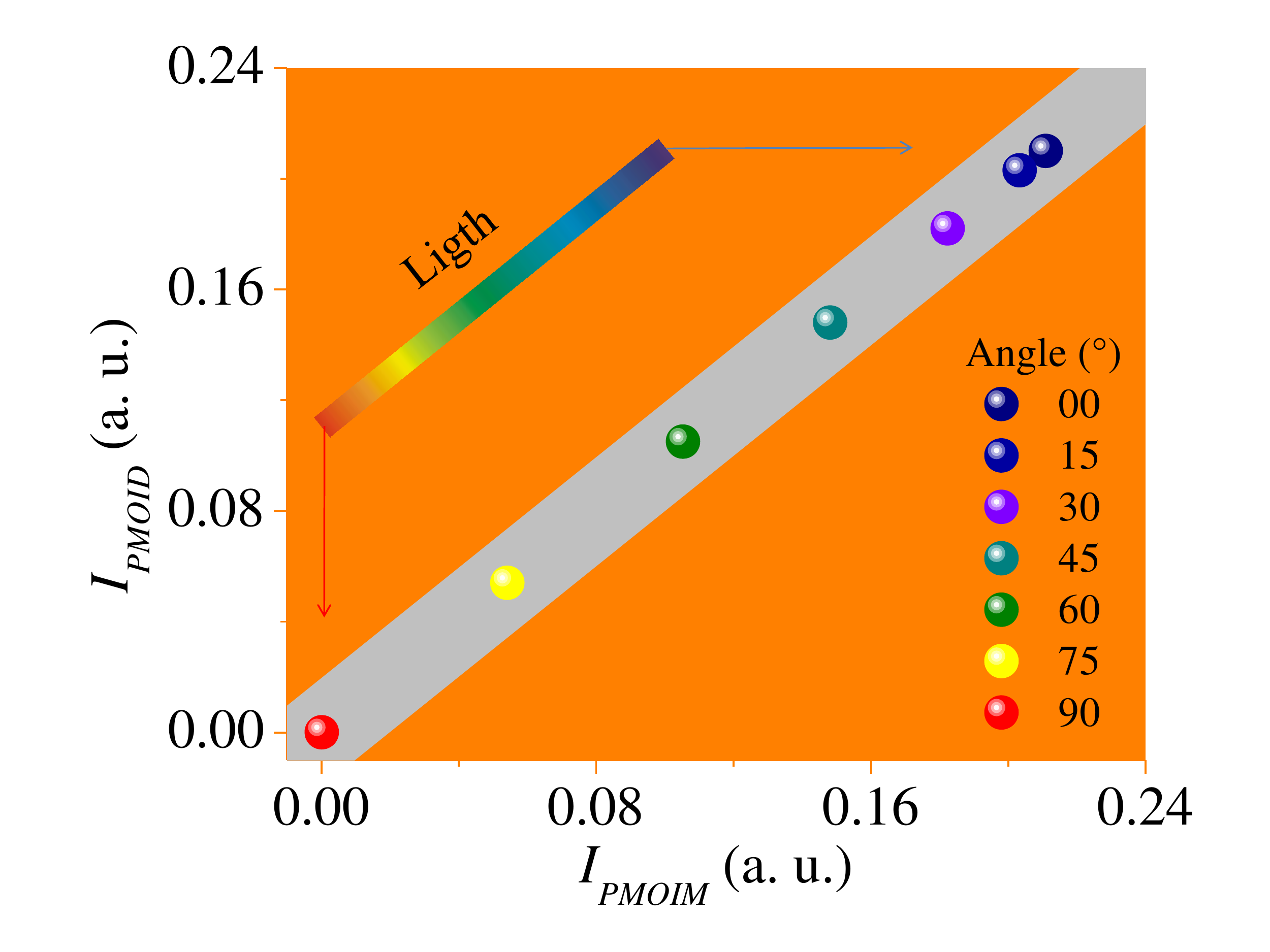}
		\caption{\label{arttype}(Color online) Magneto-optical interactions ($I_{PMOID}$ and $I_{PMOIM}$) as a function of the angle $\theta_{H}$, where we use $\Delta m_{Total} = \left(\Delta m^{D}/|\Delta m^{D}_{Max}| + \Delta m^{M}/|\Delta m^{M}_{Max}|\right)cos(\theta_{H})$ with $H_{C}^{D}$ = 3 kOe, $\Delta J_{D} = 1\% H_{C}^{D}$, $H_{C}^{M}$ = 6 kOe and $\Delta J_{M} = 1\% H_{C}^{M}$. It is shown also the best conditions to observe the maximum number of magneto-optical interactions.}
		\label{puga}
	\end{center}
\end{figure}

\section{Conclusion}

The behavior of magneto-optical interactions in magnetic structures revealed two types of predominant magnetic states, i. e., demagnetized and magnetized. Understanding how each state arises due to the different effects produced during the magnetization process makes its study of fundamental importance for applications of devices in areas such as quantum computing and engineering. Our results also represent an efficient way to describe the behavior of magneto-optical interactions by directly considering the angular dependence of the interactions undergoing the magnetic field during the magnetization process. Furthermore, the results obtained show that the effects that cause global magnetic states to arise influence the excitation of spin waves. In terms of fundamentals, the results show a significant advance in understanding the behavior of magneto-optical interactions in structures.

\section*{Acknowledgements}

This research was supported by Conselho Nacional de Desenvolvimento Cien-tífico e Tecnológico (CNPq), Coordenação de Aperfeiçoamento de Pessoal de Nível Superior (CAPES), Financiadora de Estudos e Projetos (FINEP), Centro Multiusuário de Pesquisa e Caracterização de Materiais da Universidade Federal Rural de Pernambuco (CEMUPEC-UFRPE), and Fundação de Amparo à Ciência e Tecnologia do Estado de Pernambuco (FACEPE). 

\section*{Conflicts of interest}

The authors declare no conflicts of interest.

\section*{Data availability statement}

Data underlying the results presented in this paper are not publicly available at this time but may be obtained from the authors upon reasonable request.

\bibliographystyle{MiKTeX}

\end{document}